\documentclass{ere04}

\usepackage{graphicx}    

\def\beq{\begin{equation}}
\def\eeq{\end{equation}}

\def\bea{\begin{eqnarray}}
\def\eea{\end{eqnarray}}
\def\ba{\begin{array}}                  
\def\ea{\end{array}}

\begin{document}

\title*{Stability and thermodynamics of black rings}
\author{G. Arcioni\inst{1}\and
E. Lozano-Tellechea\inst{2}}
\institute{Racah Institute of Physics, The Hebrew University of Jerusalem, 
Jerusalem 91904, Israel.
\texttt{arcionig@phys.huji.ac.il}
\and Department of Particle Physics, Weizmann Institute of Science, 
Rehovot 76100, Israel.
\texttt{ernesto.lozano@weizmann.ac.il}}
\maketitle

We study the phase diagram of $D=5$ rotating black holes and the black
rings discovered by Emparan and Reall. We address the issue of
microcanonical stability of these spacetimes and its relation to
thermodynamics by using the so-called ``Poincar\'e method'' of
stability.  We are able to show that one of the BR branches is always
unstable, with a change of stability at the point where both BR
branches meet. We study the geometry of the thermodynamic state space
(``Ruppeiner geometry'') and compute the critical exponents to check
the corresponding scaling laws. We find that, at extremality, the
system exhibits a behaviour which, formally, is very similar to that
of a second order phase transition.

\section{Introduction}

Our aim in the present talk is to report on investigations on the
properties of the phase diagram of rotating BH solutions (with just
one angular momentum $J$) of $D=5$ gravity. This scenario seems
appropriate to us due to the discovery made by Emparan and Reall of a
new BH phase: a rotating BH with event horizon topology $S^1\times
S^2$ --- what they called a ``black ring''~\cite{Emparan:2001wn}. A
crucial property of these black rings is that they exist in a region
of parameter space (the total mass $M$ and angular momentum $J$) which
overlaps with that of the spherical black hole, thus providing the
first known example of BH non-uniqueness in asymptotically flat
space. This raises questions on stability and possible transitions
between the different phases, which is what we wish to consider
here. The present contribution is based on Ref.~\cite{Arcioni:2004ww}.

\subsection*{Black Holes and Black Rings}
\label{sec:BRsAndBHs}

\begin{figure}
\center
\includegraphics[height=4cm]{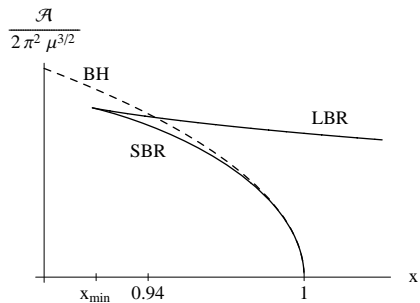}
\caption{A suitably normalised ($\mu\equiv 8GM/3\pi$) area of the
horizon of the three phases as a function of $x$. At
$x=2\sqrt{2}/3\approx 0.94$ the entropy of the large BR exceeds that of
the BH.}
\label{fig:EntropiesBHBR}       
\end{figure}

The five dimensional rota\-ting black hole~\cite{Myers:1986un} has an
upper ``Kerr bound'' in its angular momentum per unit mass, the
configuration saturating the bound being a naked singularity.  In the
following we will use the dimensionless quantity
\begin{equation} 
  x\equiv\sqrt{27\pi/32G}\, J/M^{3/2} 
\end{equation} 
as the ``control parameter''\footnote{A proper {\em order} parameter
will be defined below.} of the problem --- analogous, say, to the
temperature of a liquid-gas system in the canonical ensemble. The
rotating BH solution exists in the range of parameter space:
\begin{displaymath} 
  {\rm Black \ Hole:} \ \ \ \ \ 0\leq x<1\, , 
\end{displaymath} 
On the other hand, one can have {\em two} different black ring
spacetimes with horizon topology $S^1\times S^2$ when $x$ exceeds a
certain minimal value given by $x_{\rm min}=\sqrt{27/32}\approx 0.92$.
One of them, which we will call the ``large'' BR, can have an angular
momentum per unit mass unbounded from above. The other one, that we
call ``small'' BR, cannot. One has:
\begin{displaymath} 
  \left\{
  \begin{array}{rl} 
    {\rm Large \ Black \ Ring:} \ \ \ \ \ & 
    x_{\rm min}\leq x<\infty \, , \\
    {\rm Small \ Black \ Ring:} \ \ \ \ \ & 
    x_{\rm min}\leq x< 1\, . 
  \end{array} 
  \right.
\end{displaymath} 
The entropies are plotted in Fig.~\ref{fig:EntropiesBHBR}. Near
extremality the LBR becomes entropically favoured. This was taken
in~\cite{Emparan:2001wn} as an indication of a phase transition as we
vary $x$.

\section{Dynamical vs. Thermodynamical Stability}
\label{sec:Stability}

The first issue we want to address is that of the stability of the
classical solutions. To face this problem, a way to circumvent the
study of linear perturbations of the metric might come from
thermodynamics. In ordinary thermodynamics of extensive systems, local
{\em thermodynamical} stability (defined as the Hessian of the entropy
having no positive eigenvalues) is linked to the {\em dynamical}
stability of the system: a positive mode in the Hessian means that at
least some kind of small fluctuations within the system are
entropically favoured, implying that the system is unstable against
those. However, the simple argument~\cite{Callen} leading to this
conclusion relies on the {\em additivity} of the entropy --- a
property which does not hold for black holes (or self-gravitating
systems in general) because these cannot be thought as made up of
constituent subsystems. The best known example of this failure is the
Schwarzschild black hole: a stable configuration with negative
specific heat (i.e. a positive Hessian).

To avoid these problems, we will propose here to use the so-called
``Poincar\'e''~\cite{Poincare} or ``turning point'' method of
stability, in the way that it was first applied by Katz~\cite{Katz}
for the study self-gravitating systems. This method does not require
additivity of the entropy function, being based solely on entropy
maximisation. It was applied to BHs for the first time in~\cite{KOK}
and, contrary to the standard criteria based on the signs of the
specific heats, it does {\em not} predict any instability in e.g.~the
Schwarzschild or Kerr metrics.

We want to study the stability of solutions describing isolated BHs.
The appropriate thermodynamic ensemble is thus the microcanonical and
the relevant potential is the entropy. The crucial point is to realize
that a ``fundamental relation'' of a BH like $S=S(M,J)$ only contains
information about the {\em equilibrium} BH-series, but any discussion
based on the entropy maximum principle requires the knowledge of the
behaviour of the entropy {\em off} equilibrium. For a nonextensive
system, such behaviour will be in general different from the one given
by $S(M,J)$.  What one needs is then an ``extended entropy function'',
denoted here by $\widehat{S} = \widehat{S}(X^{\rho};M,J)$, defined
also for non-equilibrium states described by some set of
off-equilibrium variables $\{X^{\rho}\}$\footnote{They should be
thought as metric perturbations $X^{\rho}\sim \delta
g_{\mu\nu}^{(\rho)}$, since there are no other possibilities for
coordinates in configuration space.}. Equilibrium configurations are
those which obey:
\begin{equation}
  \label{EquilibriumCondition}
  \partial_\rho\widehat{S}
  \equiv\partial\widehat{S}/\partial X^\rho=0\, , 
\end{equation}
i.e. the extrema of the entropy at constant $M$ and $J$. They are
described by $X^{\rho}_{\rm eq}=X^{\rho}_{\rm eq}(M,J)$, the solutions
to Eq.~(\ref{EquilibriumCondition}) . $\widehat{S}$ gives back the
(known) equilibrium value $S$ when evaluated at $X^{\rho}_{\rm eq}$,
i.e $S=\widehat{S}(X^\rho_{\rm eq};M,J)$. A given equilibrium
configuration is {\em stable} if the eigenvalues
\begin{equation}
  \lambda_\rho (M,J) \equiv (\partial_{\rho\rho}\widehat{S})_{\rm eq}
\end{equation}
are all negative, and unstable otherwise\footnote{We take
off-equilibrium coordinates $\{X^\rho\}$ such that the matrix
$(\partial_{\rho\sigma}\widehat{S})_{\rm eq}$ is diagonal.}. In
general, the function $\widehat{S}$ is completely unknown. However,
one can infer something about its behaviour by plotting the
appropriate phase diagram along the {\em equilibrium} series.
Defining the off-equilibrium variable conjugate to $J$:
\begin{equation}
  \widehat{\omega} \equiv \partial_J\widehat{S}\, , 
\end{equation}
which reduces to $\omega\equiv\partial_JS$ when evaluated at
equilibrium (i.e. $\omega=(\widehat{\omega})_{\rm eq}$), one can show
that\footnote{Obviously, any pair of conjugate variables (like
e.g. $\beta(M)=\partial_{M}S$) will do as well.}~\cite{Katz}:
\begin{equation}
  \label{KatzIdentity}
  \partial_J\omega = (\partial_J\widehat{\omega})_{\rm eq}
   -\sum_{\rho}{(\partial_{\rho}\widehat{\omega})^2_{\rm eq}/
   \lambda_{\rho}}\, . 
\end{equation}
A change of stability takes place when one or several $\lambda_{\rho}$
approach zero and change sign. Near such a point, the term(s)
proportional to $\sim 1/\lambda_\rho$ will dominate in
Eq.~(\ref{KatzIdentity}), and therefore the equilibrium ``conjugacy
diagram'' $\omega=\omega(J)$ will exhibit a vertical
tangent\footnote{Unless also $(\partial_\rho\widehat{\omega})_{\rm
eq}=0$. In such a case one can show~\cite{Ioos} that there is a {\em
bifurcation} --- see below.}. This is often referred to as a ``turning
point''. The branch with a negative slope {\em near} a turning point
is always unstable, since for $\partial_J\omega\rightarrow -\infty$ at
least one of the vanishing $\lambda_\rho$ in~(\ref{KatzIdentity}) has
to be positive. On the other hand, if we do not know about the
stability of at least one particular point of the positive-slope (near
the turning point) branch, then we cannot say anything about its full
stability, but only that it is ``more stable'' than the negative-slope
branch (since it will have at least one unstable mode less). Also,
there might be positive eigenvalues that never change sign, and
therefore these will never show up in a conjugacy diagram.

A fundamental theorem in stability theory~\cite{Ioos} shows in fact
that the stability of a linear series of equilibria can change {\em
only} at a turning point or at a bifurcation (i.e. when a given
equilibrium series intersects with another one). Note that a change in
the sign of the specific heat takes place at a horizontal tangent, but
in nonextensive thermodynamics this has no relation to any change of
stability. 

\begin{figure}
\center
\includegraphics[height=4cm]{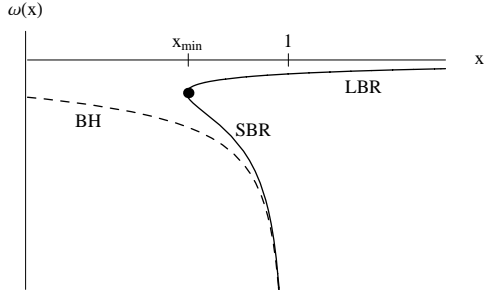}
\caption{$\omega=\partial_J S$ as a function of $x$ (i.e. $J$ at fixed
mass). There is a turning point at $x_{\rm min}$. The stability of
black rings changes there, while no changes of stability appear in the
BH branch.}
\label{fig:Stability}       
\end{figure}

The plot of the conjugacy diagram $\omega=\omega(J)$ for the BH and BR
phases is shown in Fig.~\ref{fig:Stability}. We see no turning points
(i.e. no changes of stability) along the BH branch. Given the fact
that one point in the curve (namely, the Schwarzschild limit) is
stable~\cite{Ishibashi:2003ap}, this means that the $D=5$ rotating BH
is also stable {\em unless} 1)~there are other equilibria that
bifurcate from the BH-series or 2)~there are unstable modes that this
method does not probe. As for the BRs, we see a turning point at
$x_{\rm min}$. This automatically implies that the SBR is {\em
locally} unstable. The LBR branch has to be more stable than the SBR,
although this does not prove its full dynamical stability. In fact, at
least the large-$x$ limit of the LBR is expected to suffer from
black-string-like instabilities~\cite{Emparan:2004wy}. It would be
interesting to see if such instabilities actually appear as a
bifurcation (a ``non-uniform ring branch'', say) at some value of $x$.

Although not shown in this plot, let us mention that along the BH
branch (at $x=\frac{1}{2}$) there is a change in the sign of the
specific heat at constant $J$~\cite{Arcioni:2004ww}, exactly analogous
to the $D=4$ case~\cite{Davies:1978mf}. However, this point is a
horizontal tangent in the plot of $\beta(M)=\partial_M S$, and
therefore, according to this analysis, no changes of stability are
associated to it.

\section{Thermodynamic Geometry}

It has been observed by several authors in different contexts that the
entropy function induces a natural metric on the thermodynamic state
space, and that the geometric invariants constructed out of it provide
information about the phases of the model under consideration. This
formalism was pioneered by Ruppeiner~\cite{Ruppeiner}. The Ruppeiner
metric is defined by (minus) the second derivatives of the entropy.
In our off-equilibrium formalism, where we have
$\widehat{S}=\widehat{S}(X^\rho;M,J)\equiv\widehat{S}(X^{\rho};\mu^i)$, 
this means:
\begin{equation}
  \label{RuppeinerMetric}
  g_{ij}=-(\partial_{ij}\widehat{S})_{\rm eq}\, , \ \ \ \ \ 
  g_{i\rho}=-(\partial_{i\rho}\widehat{S})_{\rm eq}\, , \ \ \ \ \ 
  g_{\rho\rho}=-\lambda_\rho\, .
\end{equation}
Using the standard techniques of thermodynamic fluctuation
theory~\cite{Ruppeiner,LandauLifshitz}, from the quadratic expansion
of $\widehat{S}$ one can compute the second moments of correlations of
any quantities around equilibrium. These turn out to be given by the
inverse elements of the Ruppeiner metric. In particular:
\begin{equation}
  \label{InverseRuppeiner}
  \langle\Delta\mu^i\Delta\mu^j\rangle = g^{ij}\, . 
\end{equation}

A most important quantity in Ruppeiner theory is the curvature scalar
$R$ of this metric. Among other properties, calculations
show~\cite{Ruppeiner} that $R$ diverges at a critical point, and that
it does so in the same way as the correlation volume:
\begin{equation}
  \label{RuppAndCL}
  R_{\rm crit} \sim \xi^d\, ,
\end{equation}
where $\xi$ is the correlation length and $d$ the effective dimension
of the system. 

The Ruppeiner metric as defined above cannot be computed explicitly,
since in general we have no knowledge about the off-equilibrium
entropy function $\widehat{S}$. However, it can be
shown~\cite{Arcioni:2004ww} that near $x_{\rm min}$ (turning point) in
the LBR branch, and near $x=1$ (vertical asymptote) in the BH/SBR
branches, the quadratic fluctuations in $\beta$ and $\omega$ (the
variables conjugate to $M$ and $J$) diverge and, furthermore, that
they can be well approximated by the {\em equilibrium} Hessian (which
in fact does diverge at those points --- see~\cite{Arcioni:2004ww} for
the explicit computation). Using~(\ref{InverseRuppeiner}) it can be
shown that:
\begin{equation}
  \label{ApproximateRuppeiner}
  g^{ij}\approx \pm \partial_{ij} S\, ,
\end{equation}
where the plus sign stands for the LBR branch near $x_{\rm min}$ while
the minus sign stands for the BH and SBR branches near $x=1$. That is,
{\em near} $x_{\rm min}$ and $x=1$, the ``effective'' elements of the
Ruppeiner metric are just a few and, moreover, they can be computed
from the equilibrium entropy.

\begin{figure}
\center
\includegraphics[height=4cm]{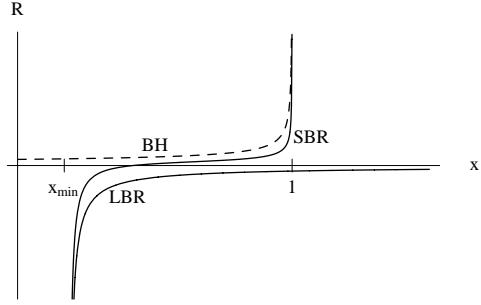}
\caption{Thermodynamic curvature scalar. The relevant 
values are those near $x=1$ and $x_{\rm min}$.}
\label{fig:Ruppeiner}       
\end{figure}

We have computed the thermodynamic curvature for the BH/BR system and
the result is plotted in Fig.~\ref{fig:Ruppeiner}. For convenience,
the plot shows the values of $R$ as computed
from~(\ref{ApproximateRuppeiner}) (with the plus sign) for all $x$.
One can see that not only the metric, but also the thermodynamic
curvature scalar diverges at $x_{\rm min}$ and $x=1$.

\section{Critical Exponents}

Finally, we turn now to the study of the scaling relations familiar
from RG-theory in Statistical Mechanics at the points $x_{\rm min}$
and $x=1$. We first define the appropriate susceptibilities and
critical exponents that are suited to the microcanonical
ensemble~\cite{KaburakiSL}:
\begin{equation} 
    \chi_J  \equiv  
    \partial_M^2S \sim  
      \epsilon_M^{-\alpha}\, , \   
      \epsilon_J^{-\varphi}\, , \ \ \ \ \ 
    \chi_M  \equiv  
    \partial_J^2S \sim  
      \epsilon_M^{-\gamma}\, , \    
      \epsilon_J^{-\sigma}\, .  
\end{equation} 
These are the natural generalisations of the expressions familiar from
other ensembles. At $x_{\rm min}$ and $x=1$ the parameters
$\epsilon_M$ and $\epsilon_J$ are defined to be, respectively:
\begin{equation} 
    \epsilon_M  =  
    \textstyle\frac{M_{\rm min}-M}{M_{\rm min}}\, , \ 
    \epsilon_M  =  
    \textstyle\frac{M-M_{\rm ext}}{M_{\rm ext}}\, ; \ \ \ \ \ \ 
    \epsilon_J  =  
    \textstyle\frac{J-J_{\rm min}}{J_{\rm min}}\, , \  
    \epsilon_J  =  
    \textstyle\frac{J_{\rm ext}-J}{J_{\rm ext}}\, .  
\end{equation} 
Other critical exponents tell us about the behaviour of the order
parameter $\eta$ of the transition considered. In general, there is no
obvious choice for such order parameter. Kaburaki proposed
in~\cite{KaburakiSL} to choose, as an order parameter in BH phase
transitions, the difference of the conjugate variables between the two
phases. In our case this means:
\begin{equation} 
  {\rm At \ } x_{\rm min} {\rm :}  \ \ 
  \eta \equiv \omega_{\rm SBR} - \omega_{\rm LBR}\, . \ \ \ \ \ \ \ 
  {\rm At \ } x=1 {\rm :} \ \
  \eta \equiv \omega_{\rm SBR} - \omega_{\rm BH}\, .
\end{equation} 
With these definitions, the critical exponents $\beta$ and $\delta$
familiar from statistical mechanics are given by:
\begin{equation} 
  \eta \sim \epsilon_{M}^{\beta} \sim \epsilon_{J}^{\delta^{-1}}\, . 
\end{equation}
All these critical exponents are readily computed to be:
\begin{equation} 
  {\rm At \ } x_{\rm min} {\rm :} \ \ 
  \alpha, \varphi, \gamma, \sigma = \textstyle\frac{1}{2}\, ; \  
  \beta, \delta^{-1} = \textstyle\frac{1}{2}\, . \ \ \ \ \ \ \ 
  {\rm At \ } x=1 {\rm :} \ \ 
  \alpha, \varphi, \gamma, \sigma = \textstyle\frac{3}{2}\, ; \  
  \beta, \delta^{-1} = -\textstyle\frac{1}{2}\, .
\end{equation} 
The scaling relations involving the critical exponents defined so far
are given by~\cite{LandauLifshitz}:
\begin{equation}
    \alpha + 2\beta + \gamma  =  2\, , \ \ \ \ \  
    \beta(\delta - 1)  =  \gamma\, , \ \ \ \ \ 
    \varphi(\beta+\gamma)  =  \alpha\, .   
\end{equation}
One can check that these scalings are satisfied both at $x=x_{\rm
min}$ and at $x=1$.

However, at a critical point of a second order phase transition further
scaling laws involving other critical exponents are satisfied. The
remaining critical exponents include those related to the behaviour of
the correlation length, whose divergence at a critical point is
expressed in terms of the exponents $\nu$ and $\mu$ as:
\begin{equation}
  \xi \sim \epsilon_M^{-\nu} \sim \epsilon_J^{-\mu}\, ,
\end{equation}
which obey the scaling laws~\cite{LandauLifshitz}:
\begin{equation}
  \label{Scalings2}
  2-\alpha  =  \nu d\, , \ \ \ \ \ \  
  \mu(\beta+\gamma)  =  \nu\,. 
\end{equation}
Even if there is no obvious way to compute (or even define) $\xi$ for
a BH geometry, Eq.~(\ref{RuppAndCL}) provides us with an explicit
algorithm to compute $\nu d$ and $\mu d$ once we have computed the
divergence of the thermodynamic curvature scalar. The result
is~\cite{Arcioni:2004ww}:
\begin{equation} 
  {\rm At \ } x_{\rm min} {\rm :}  \ \  
  \nu d = \mu d = 1\, . \ \ \ \ \ \ \ \ 
  {\rm At \ } x=1 {\rm :}  \ \  
  \nu d = \mu d = \textstyle\frac{1}{2}\, .
\end{equation} 
Relations~(\ref{Scalings2}) are obeyed at $x=1$ for any effective $d$,
but they are not satisfied at $x_{\rm min}$.

\section{Conclusions}

Using the Poincar\'e method of stability we have shown that the SBR is
unstable not only globally (see Fig.~\ref{fig:EntropiesBHBR}) but also
locally. Stability changes at $x_{\rm min}$, where
Fig.~\ref{fig:Stability} exhibits a turning point.  Near $x=1$ the BH
and SBR branches behave in a way similar to that of a system near a
critical point. However, we cannot speak about a second order phase
transition here, since at least one phase (the SBR) is unstable. Also,
we do not know about any phases beyond the ``critical point''. The
found scaling behaviour has to be interpreted then as describing,
formally, the properties of the system {\em only} along the stable
directions in configuration space. However, we find such a property
nontrivial. Finally, nothing special happens at the point in the BH
branch where the specific heat changes sign ($x=\frac{1}{2}$) or at
$x=2\sqrt{2}/3\approx 0.94$ (see Fig.~\ref{fig:EntropiesBHBR}).  Given
this fact, {\em if} both the BH and LBR are stable around $x\approx
0.94$, the only possibility we are left with is that of a first order
phase transition between both. It is also interesting how the
thermodynamic curvature seems to reproduce the behaviour of the
correlation length~$\xi$. However, without an explicit computation of
$\xi$ for these BH geometries, Eq.~(\ref{RuppAndCL}) is just an Ansatz
for it. It would be desirable to understand this in a better way.

\section*{Acknowledgments}

E.L.-T. wants to thank the organizers of the XXVII Spanish Relativity
Meeting for a successful and interesting conference and for their kind
invitation to give this talk, as well as to the participants J.
L. F. Barb\'on, D. Cremades, R. Emparan, R. Myers and J. Russo for
their interesting questions and comments on the present work.  This
work has been supported in part by a Marie Curie Fellowship under
contract MEIF-CT-2003-502349 and the Spanish Grant BFM2003-01090.

\end{document}